# Using Computational Essays to Scaffold Professional Physics Practice

Tor Ole Odden and Anders Malthe-Sørenssen
Center for Computing in Science Education, Department of Physics,
University of Oslo, Norway

**Abstract:**
This article describes a curricular innovation designed to help students experience authentic physics inquiry with an emphasis on computational modeling and scientific communication. The educational design centers on a new type of assignment called a computational essay, which was developed and implemented over the course of two semesters of an intermediate electricity and magnetism course at the University of Oslo, Norway. We describe the motivation, learning goals, and scaffolds used in the computational essay project, with the intention that other educators will be able to replicate and adapt our design. We also report on initial findings from this implementation, including key features of student-written computational essays, student reflections on the inquiry process, and self-reported conceptual and attitudinal development. Based on these findings, we argue that computational essays can serve a key role in introducing students to open-ended, inquiry-based work and setting the foundation for future computational research and studies.

1. Introduction

The way we teach physics often bears little resemblance to the way we do physics. Classical physics education tends to focus on the twin pillars of theory and experiment. In practice, this mostly entails students re-deriving existing knowledge through problem-solving, derivations, and structured labs. However, there is much more to physics than just theory and experiment; in recent decades, for example, computation has been added as a "3rd pillar" of physics[1,2]. Despite this development, many physics departments have been slow to bring computation into their educational programs[3,4]. Professional physicists also work within the scientific community, which supports them in continually communicating, reviewing, and refining results to ensure their robustness. However, standard physics courses include few opportunities to practice this kind of communication, refinement, or community engagement, except in limited circumstances such as innovative laboratory courses[5].

This disconnect between education and professional practice is unfortunate. If we actually intend to train students to "think like physicists," we must go beyond basic exercises and help them understand what it really means to work within the discipline. We must equip them with examples of how physics can be used to explain, understand, and explore nature, how it provides robust approaches that are widely applicable, and how the discipline of physics is a creative and open enterprise that favors exploration, not just replication.

At the University of Oslo, we have created an educational design that aims to address this gap. Our design focuses on helping students simultaneously gain an initial experience in doing a computational physics investigation and learn the basics of communicating their methods and results. The core of this design is a new type of assignment called a computational essay.

In this article, we describe our computational essay educational design, with the goal that others will be able to replicate it. Specifically, we describe

1. Our motivation, philosophy, and the educational context that allows for this kind of curricular innovation
2. Our goals for the project, and the role that it fulfills within our physics curriculum
3. Some initial results from a pilot study and full implementation in a 3rd-semester electricity and magnetism course

**2. Motivation and Philosophy**

In this section, we describe the motivation behind our work—the theoretical foundations we build it on, the need for these curricular changes, and our primary innovation, the computational essay.

*2.1 The need for more authentic scientific practice in education*
The theory underlying this project is a so-called *practice perspective of learning*, which has seen increasing use in research in physics education (6,7) and science education (8,9). This theory posits that, rather than viewing science learning as a process of acquiring increasingly correct and sophisticated facts, mental models, or techniques for logical deduction, we can instead view it as a process of acquiring a set of "practices" from a community (8)—in this case, the physics community. At the pre-college level, this often means positioning students such that they are able to question, test, and push back on scientific ideas, and engage in authentic scientific behaviors like argumentation (9–12). At the college level and above, this shift entails a move towards increased disciplinary authenticity—that is, having students focus more on "doing science" than "learning about science."

This shift has required educators to reexamine many of the fundamental assumptions underlying how physics has been traditionally taught. For example, if one's goal is to have students develop and defend physical models, one may have to adopt new technologies and teaching methods, such as computational modeling (6,13). One may also have to rethink the tradition of having students work through sets of discrete physics problems with only a single right answer, and instead focus on the iterative development and refinement of causal models (14). If one is aiming to incorporate communication, writing, or peer review, one may have to give students tasks that are open-ended or unstructured in order to give them "something to write about" (15,16). This is even the case in more experimental contexts, such as physics labs, which are meant to give students a taste of what it means to "do science" rather than just "learn about science" (17,18). Without careful thought, such contexts can easily slip into an overly-guided "cookbook"-style format, leaving little room for authentic communication, argumentation, or explanation.

Since the fall of 2018, we have been in the process of revising our physics curriculum to incorporate this more practice-oriented approach to physics teaching and learning. This revision is based around a new kind of assignment specifically developed for these purposes, called a computational essay.

*2.2 Computational Essays in Physics*

Computational essays are a genre of scientific writing that combines live code, written prose, mathematics, and pictures or diagrams in order to make an argument, explain an idea, or tell a story. An example computational essay, written by a pair of students from the present study, is shown in Figure 1.

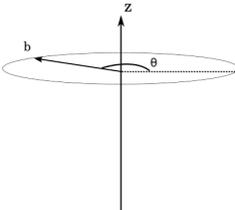

**Figure 1:** Example computational essay, written by a pair of students in the present study. The essay is modeling the motion of a high-energy charged particle within a magnetic bottle.

The term "computational essay" was originally coined by diSessa (19), and later taken up by Wolfram (20) who described them as "an intellectual story told through a collaboration between a human author and a computer." In this study, we conceptualize computational essays as a hybrid between an open-ended computational project and an essay-like writing assignment.

Computational essays are usually written in computational *notebooks*, development environment that allow one to intersperse blocks of live, executable code with embedded text, equations, pictures, or animations.[1] Many readers will likely be familiar with Mathematica, a brand of mathematical notebooks that has been available under license for decades (21). More recently, Jupyter notebooks have emerged as an open-source alternative that allow for programming in widely-used scientific computing languages such as Python and R (22).

Computational essays have become an increasingly common tool for sharing preliminary scientific results, data analysis methods, or as supplements to scientific papers, both in physics (23) and in other computation-heavy fields such as data science (24). Notably, the blocks of code in a notebook can be executed independently of one another, or re-executed independent of the rest of the script. For this reason, they are also useful tools for computational exploration and sharing works-in-progress within research teams (25).

We became interested in the educational possibilities of computational essays for several reasons. First, as previously mentioned, they are an increasingly-used part of professional scientific practice—in fact, both authors of this paper frequently use them in their own professional practice for both exploration and communication. Thus, we saw this as a way to bring our curriculum more in line with how professional physicists and data scientists work and share results. There have also been suggestions that notebooks are more useful for communicating certain kinds of ideas and results than standard scientific papers (26). For this reason, we feel that it is important to train our students in this emerging genre to make sure they are prepared for the job market.

More importantly, we see computational essays as an ideal way to bring in more room for open-ended, creative, and communicative work into the physics curriculum. In this regard, computational essays may help solve several thorny, recurring problems that frequently crop up when introducing open-ended project-based learning into physics education. For example, one issue that always comes up when students do open-ended projects is "what will they turn in?" Computational essays, by their nature, solve this problem, since the essay is both the computational simulation and the end-product, and in writing such an essay, students are given a chance to experience authentic scientific communication practices. Another problem is the difficulty of documenting students' inquiry processes—that is, it is often difficult to see how a student has actually gone about doing an open-ended project. Computational essays solve this issue since (when properly written) they communicate a complete process of inquiry: a research question, computational methods used to answer it, and conclusions. A third issue is how to support students in pursuing research questions that they themselves find interesting, rather than requiring them to work within a (usually limited) set of pre-defined questions. Because computational essays are based on computational simulations, they allow for a huge variety of different types of research questions, constrained only by students' coding ability, knowledge of numerical methods, and imagination.

3. **Description of computational essay design**

---

[1] To be clear, here we are considering computational essays to be a genre of writing, which is built on top of computational notebooks. However, there are other ways of writing computational essays that do not use notebooks; for instance, one can embed interactive programs into a website. One prominent example of this style of essay the work of Rhett Allain for *Wired*(37).

*3.1 Institutional context*

The context of our study is the physics department at the University of Oslo (UiO), a large research-focused university in Norway with an international presence. The University of Oslo physics program features a 3-year bachelor's degree with a heavy emphasis on scientific programming (27). Introductory physics courses tend to be relatively large, enrolling 200-300 students, and serve several majors including physics, astronomy, materials science, electrical engineering, geoscience, and physics teacher preparation.

Our intervention was developed within a 3$^{rd}$-semester electricity and magnetism course, taught by the second author. We chose to target this course because by the 3$^{rd}$ semester, most physics students at the University of Oslo have taken a programming course, a numerical methods course, have a baseline of mathematical preparation, and have attained some familiarity with computational modeling during their previous two semesters. However, they have not had significant training in any forms of scientific communication, nor have they had a chance to pursue any type of open-ended computational investigations.

*3.2 General parameters of our intervention*

Our goal of our educational design was to have students define a research question and perform an open-ended computational investigation using the physics they had learned thus far in their course. Then, they would write up their results in a computational notebook, and present these results to their peers in a mock research-group meeting.

We implemented the design in two phases: first, we ran a pilot implementation in the fall semester of 2018. This pilot was voluntary; students had the choice to either write and present a computational essay or present a report-style oral presentation on a chosen physics topic (a standard part of the course prior to that semester). 17 students chose the computational essay option, working singly or in pairs to produce 11 computational essays. Building on this pilot, during the Fall semester of 2019 we expanded this design to be a mandatory part of the course, requiring all students to write and present computational essays either individually or in pairs.

Our design incorporates several explicit learning goals:

1. Learning goal 1: **Students will identify an interesting, tractable question for their investigation and answer it using a computational simulation**. Students will identify a real-world system that *can* be analyzed using the principles they've learned; Students will find physical values from that system and set the parameters in their simulation to match those values; Students will implement some aspects of the physics they have learned during the semester within their computational simulation

2. Learning goal 2: **Students will identify assumptions inherent to the computational model, and modify it to address some of those assumptions**. Students will identify assumptions inherent to the simulation of their chosen system; Students will determine which assumptions can be addressed within the simulation; Students will modify their simulation to address one or more of these assumptions

3. Learning goal 3: **Students will either program a computational simulation from scratch or build on example code and revise/augment it to answer their question**. Students will effectively use basic computational elements such as variables, loops, functions, and classes; Students will use basic modeling techniques like numerical integration methods; If building off example code, students will comprehend, revise, and modify that code to account for assumptions and values based on the physical system they are analyzing; Students will write programs that run without errors

4. Learning goal 4: **Students will effectively communicate the methods and results of their investigations**. Students will clearly state the questions they have chosen, and explain why they are important or interesting; Students will communicate the trajectory of their investigations within the body of the computational essays; Students will use good coding practices, logically group together code, and communicate the meaning of each code chunk; Students will evaluate and communicate the meaning of their code outputs and relate it back to their original question; Students will use simulation results to form an argument that answers their original question in a discussion or conclusion section

In order to help our students be successful with their projects, we provided them with several forms of scaffolding. The first scaffold was a written assignment description which outlined the motivation behind the project, framing it as an opportunity for students to try their hand at a more authentic scientific activity. In this description, we called out the ways actual scientists use computational simulations to define research questions, answer those questions, and communicate their results to one another. We also laid out our objectives for the procedure they were expected to follow in developing their projects, the final product they were expected to turn in, and the time commitment they could expect to make. This document additionally provided an overview of the supports they could draw on, such as example computational essays they could consult, help-sessions that would be provided, and the fact that they were encouraged to complete the assignment with a partner.

Appended to the project description was a copy of the rubric we planned to use in grading the final computational essays. This rubric had been developed based on the learning goals described above, and featured 5 primary categories, each equally weighted and evaluated on a scale of 0-4. At the top end of the scale (4 points), the categories and descriptions were as follows:

1. **Investigation question:** There is an investigation question, it is physically meaningful, and it requires significant additions to the example simulation to answer
2. **Coding:** The code works and there are significant additions to the code (i.e., several new steps or blocks added to the simulation)
3. **Physics in the simulation:** Physics principles have been used to augment the simulation, and it is clear how they were derived and applied in the code
4. **Conclusions:** There is a conclusion which describes the results, interprets their meaning, uses them to answer the original question, and justifies their reasonability

5. **Written report:** There is a report which clearly explains the steps of the investigation, fleshed out with at least 1 picture or diagram (beyond the pictures/diagrams given in the original simulation)

In the project description, we specified that the students would need to score at least 70% (14/20 possible points) on this rubric in order to pass. Although these guidelines were relatively lenient, by providing them to the students we aimed to establish a desired minimum threshold for their performance, with the hope that students would generally exceed this threshold. In practice, we found that most students did.

Because the students were constrained in the amount of time they were expected to put in (roughly 10 hours), and because many students were less-than-comfortable with programming, the physics covered in the course, or both, we provided the students with a number of pre-made simulations that were intended to act as "seeds" for the students to build out into fully-fledged computational essays. These "seed" programs were stripped-down simulations of physics phenomena, written in Jupyter notebooks, that ran without errors and provided an overview of the basic theory behind the phenomena. However, we specifically designed the simulations such that they did not illustrate any especially interesting results. In the first iteration of the project, these "seeds" consisted of simulations of a storm cloud, lightning bolt (in 2D), cyclotron, railgun, and magnetic bottle (28). In the second iteration we added a simulation of a neuron, polar molecules in a liquid, and an electric dipole antenna. However, we also encouraged students to build off of other computational assignments from the course, and/or write their own simulations from scratch if they felt inspired to do so.

Our intention behind these simulations was three-fold. First, we intended to provide students with a fruitful starting point, to mitigate the risk that they would spend excessive amounts of time struggling with how to start. This choice was based on our understanding that finding an interesting problem to investigate is often the most difficult part of the research process, so we aimed to inspire students by providing them with simulations they could explore and play around with. We felt this was authentic to professional scientific practice, since in research contexts one seldom starts new computational project entirely from scratch. Second, by placing these examples in a Jupyter notebook, we provided students some initial contact with notebooks and nudged them towards using notebooks for their programming environment, rather than the more standard IDEs they had used previously. Third, these simulations allowed us to provide students with some explicit suggestions for questions they could pursue in case they were having difficulty finding interesting questions on their own. These suggestions took the form of prompts at the end of the notebooks such as, "what are the effects of special relativity on the cyclotron simulation?" or "A reversed magnetic bottle is known as a 'bionic cusp'; how does a particle's behavior in a bionic cusp differ from that in a magnetic bottle?"

In order to both concretize our expectations and provide the students with additional inspiration, we also furnished them with examples of fully-written computational essays. These essays were explicitly designed to showcase best practices that we intended our students to adopt—for example, a strong investigative narrative, judicious use of images and diagrams, iterative development of a computational model, use of functions, code commenting and documentation, cited sources, reasonability checks, and explicit discussion of model limitations.

The essays are available at the University of Oslo's Online Computational Essay Showroom (29). During the pilot implementation of the project, we only provided students with a single example, answering the question "how much current would be needed to use a railgun to resupply the International Space Station?" After the pilot, we added several more essays written by faculty, staff, and hired assistants, as well as several student-written essays collected during the initial phase of the project (and posted with permission).

During the periods when students were working on their computational essays, we additionally provided weekly drop-in help sessions, where students could pose questions about theory, code, or logistics related to the project. During the initial implementation these were staffed by the first author, while in the subsequent implementation they were staffed by hired teaching assistants.

Finally, in order to pass the project, students were required to present their results to their peers in mock research group meetings. Each "meeting" featured 5-9 presentations; students were given 8 minutes to present if they had worked individually or 13 minutes if they were working in a pair, plus two minutes for questions. During these presentations, the students were asked to project their completed notebooks onto a screen and use them as the primary presentation aid, thereby necessitating that they showcase their code to their peers. These presentations were intended to function both as an opportunity to develop communication skills and as a motivator for the students, in that they were required show their work to their peers, not just to their grader.

## 4. Learning outcomes and evaluation

In this section, we describe the results of the project in the form of learning outcomes and student reactions.

Since this was an open-ended educational design, it did not lend itself to directly assessing student learning—that is to say, it is entirely possible that students came into the project with all of the content knowledge, coding, and presentation skills necessary to complete it. However, we suspect that this is unlikely to be the case; to our knowledge, prior to their participation in the computational essay project, most students had never done an open-ended, physics-based computational project—at least not as part of their coursework. As a check on this assumption, and in order to understand the students' experiences with the computational essay writing process, we interviewed a number of the students and asked them to describe to us the process they went through in writing their essays. Specifically, we asked about their source of inspiration for their topic, key challenges and lessons learned, and concepts or techniques that they felt more comfortable with as a result of having done the project. Additionally, we asked interviewees several more general questions about their computational background, their views on the connection between computation and physics learning, and their reflections on the role of creativity within their projects. Interviews were conducted in English, Norwegian, or a mixture of the two (depending on the preferences of the interviewees) and interviewees were incentivized with gift cards.

During the pilot phase of the project, we collected 10 of these interviews (capturing all participating students except one pair who declined to be interviewed). During the second iteration, we collected 12 more interviews, using the same interview protocol. We also collected computational essays from consenting students during both implementations, resulting in 11 collected essays during the pilot phase and 58 essays during the second iteration. Although the number of students enrolled in the course during the second iteration (on the order of 200) substantially outnumbers the essays we collected, the number of collected essays reflects the fact that not all students consented, and many consenting students worked in pairs.

In what follows, we provide an overview of these results in the form of a general description of the essay corpus, followed by excerpts from the interviews. Here, we are not aiming for a complete accounting of everything the students learned in the process. Doing so would require additional measures that we did not use, such as pre- and post-tests. Additionally, as previously mentioned, our goals with the project were fairly wide-ranging: we wanted students to not only improve their understanding of physics, but also gain exposure to authentic computational physics practices in a limited, scaffolded environment; practice scientific communication; and get a "taste" of the scientific inquiry process. So, below we present a subset of results intended to display key themes from the corpus as well as to illustrate the variety of essays and responses, with the intention that readers will be able to see what is possible and judge how well it fits with their own course goals. Where appropriate, we highlight challenges encountered and aspects we intend to change in future iterations of the project.

We have broken these results into three sections:

1. *Overview of computational essay corpus.* This is an overview of the general features of the computational essays collected across the two iterations of the project—topics chosen by the students sorted by "seed" programs, trends in essay structure, and general level of programming sophistication on display

2. *Student reflections on the computational essay writing process.* This is an overview of student responses and reflections on the process of writing computational essays— what they found easiest and most challenging, trends in how they approached the task, and their overall takeaways from the project

3. *Conceptual learning and attitudinal development.* This is an overview of student reflections on both the physics understanding they gained in the process and the ways the project affected their views and attitudes towards the discipline of physics

*4.1 Overview of computational essay corpus*

We begin with an overview of the computational essays collected from iterations 1 and 2 of the project. A list of the topics students chose to address is shown in Table 1, broken up based on which "seed" program the students built off of (if any). In cases where multiple students or groups chose substantially similar topics, we have noted this with a number after the topic. We have also grouped together essays that focus on similar approaches, such as applications, comparisons, or explorations of the targeted system.

| Base Simulation | Computational Essay Topics |
|---|---|
| Railgun | - Applications of a railgun for…<br>  - Sending a satellite into orbit around the sun (2)<br>  - Accelerating aircraft to high velocities (2)<br>  - Powering a train line (2)<br>  - Getting a student to class on time<br>  - Launching a person at 1 m/s<br>  - Sending a person to the top of a mountain<br>  - Propelling a space elevator<br>  - Launching a Tesla Model S car to Mars<br>  - Resupplying an offshore oil platform<br>- Comparison of a railgun to…<br>  - An ordinary pirate cannon<br>  - A Tsar bomb when used for orbital bombardment<br>- Optimizing the mass/angle of a railgun shot to maximize distance (2)<br>- An account of all of the limitations with the example railgun simulation |
| Cyclotron | - Applications of a cyclotron for…<br>  - Cancer treatment (2)<br>  - Communication purposes by modulating the magnetic field<br>  - Driving nuclear reactions<br>- Exploration of the effects of…<br>  - Special relativity in a cyclotron (3)<br>  - Varying voltage functions plus special relativity (2)<br>  - An electron in a cyclotron<br>- Comparison of a cyclotron and a synchrocyclotron when accelerating particles to relativistic speeds (2) |
| 2D-Lightning | - Exploration of the safety of…<br>  - Being in a car during a lightning strike (3)<br>  - Body positions during a lightning strike<br>  - Locations on the ground during a lightning strike<br>- Exploration of lightning behavior in the presence of…<br>  - A lightning rod on a building<br>  - The Eiffel Tower<br>  - Lightning rods with different lengths, thickness, or compositions (2)<br>- Simulating the stepped leaders that cause lightning strikes |
| Stormcloud | - Exploring the effects of…<br>  - Cloud materials with different electric permittivity<br>  - Dimensions of a storm cloud<br>- Simulating how storm clouds generate lightning (2) |
| Magnetic bottle | - Comparison of a magnetic bottle to a…<br>  - Biconic cusp (2)<br>  - Magnetic mirror |

|  | - Trapping highly energetic particles in a magnetic bottle
- Trapping a particle using multiple magnetic bottles
- Testing the conditions under which a particle escapes a magnetic bottle
- Simulating multiple interacting particles within a magnetic bottle |
|---|---|
| Nerve Cell | - Simulating a nerve cell, with the addition of vicious friction on the moving ions (2)
- Exploration of membrane potential in a re-polarizing neuron |
| Dipole molecule | - Orientation of water molecules in the presence of a positive charge
- Simulating polar water molecules in 3D rather than 2D |
| Other (diverse topics) | - Simulating relativistic particles in the LHC
- Using an electric field do you need to break up the quarks in a neural meson
- Using a Helmholtz coils for on-site radiation shielding
- Accelerating spaceships by shooting them through a hole in the moon
- Accelerating a charged cannonball using a linear particle accelerator
- Using the weight force be used to warm up a room through EM-induction
- The effects of inductive coil on a circuit used for EM-braking
- Car lane control using magnetic fields
- Simulating the Northern lights
- Simulation of a roller-coaster using EM-effects for acceleration and braking
- The use and efficiency of induction heaters vs. resistive heaters
- Thermal energy produced by an induction stove
- Simulating a cathode-ray tube for use in TVs
- Charging electric cars as they drive |

**Table 1:** Student computational essay topics, separated by "seed" program. In cases where multiple students or groups addressed a similar topic, we have added the number of essays to the end of the topic. We have also grouped together essays that focus on similar themes, such as applications, comparisons, or explorations.

As one can see from Table 1, there is a great variety of topics chosen. However, there are also certain commonly-addressed topics; for the purposes of illustration, we present three snapshots of examples, in Figures 2-4. Figure 2 shows a written introduction from an essay that analyzed the heat output and distribution from induction cookstoves; this is an example of an essay on a "diverse topic" (not based on any of the provided "seed" programs). Figure 3 shows a code snippet from an essay that simulated the use of railguns to power a space elevator; it is exemplary of the many essays on diverse applications of railguns. Figure 4 shows a graph from an essay comparing relativistic and non-relativistic particles in a cyclotron. This was an especially popular topic, addressed both by itself and as a component of several other analyses (such as the use of cyclotrons for nuclear power generation). We suspect this popularity stems in part from the fact that many students were co-enrolled in an astrophysics course that covered special relativity; however, this topic was also one of the prompts provided to students at the bottom of the cyclotron "seed" simulation.

> ## 1.1 Abstract
>
> *This computational essay attempts to find the electromagnetic field and thermal power output that is generated and distributed in an induction cooktop. Starting with a copper coil, a magnetic field is created by applying a current through the coil. Copper is used due to its great conductal properties. The change in the magnetic field induces an electric field in the bottom material of the cooking pot. This pot is made of cast iron. The charges move about in a circular motion, consistent with the hole in the coil centre below it. The resistance in the cast iron then yields thermal energy from the charges kinetic energy. This thermal heat is, as the electric field induced, located above the periphery of the coil centre hole. The heat from there will spread through the pot's bottom. An improvement to this simulation would be to take the AC frequency into account when finding the magnetic field. Moreover, it would also be possible to find the convectional properties of the heat transfer in the pot, and through its contents - finding the temperature as a function of time.*
>
> ## 1.2 Introduction
>
> *This computational essay aims to investigate the electromagnetic physics and thermal power output is generated and distributed in an ordinary induction cooktop. The most important component of such a device is a coiled wire of some conductive material. Other components includes power lines, cooling fans, thermal sensors and many more, but in this essay we are placing our focus on the conductive coil. By leading a current through the coil, w, a magnetic field is induced as shown from the right hand rule. This magnetic field will cause electrons in the pot bottom to move in a circular motion. The resistance in this material will then cause an energy release from the kinetic energy of the moving electrons to thermic energy from the materials resistance. This thermic energy will then heat the bottom of the pot.*

**Figure 2:** Introduction of a computational essay on the thermal power output and distribution produced by an induction cooktop

```python
class SpaceElevator:
    """ Class representing a space elevator on the moon or on the earth.
    """
    
    def __init__(self, x0=1, L=1.4, r_cable=0.05, m=1, dt=0.01, moon=False):
        self.x0 = x0  # initial position above surface [m]
        self.L = L  # distance between cables [m]
        self.r0 = 6371000  # radius of earth [m]
        self.m_E = 5.972e24  # mass of earth [kg]
        self.m_M = 7.348e22  # mass of moon [kg]
        self.r_cable = r_cable  # radius of cables [m]
        self.GEO = 35786000  # radius of Earth GEO orbit [m]
        self.moon_surface_E = 386139100  # moon opposite surface dist from Earth center [m]
        self.moon_surface_M = 1737100  # radius of moon
        self.L2_E = 448900000  # Earth/Moon L2 position from Earth center
        self.L2_M = self.L2_E - self.moon_surface_E
        self.G = constants.G  # Newtonian constant of gravitation
        self.mu0 = constants.mu_0  # Magnetic premeability vacuum
        self.moon = moon  # Check to calculate for elevator on moon
        self.ln = np.log(L+(r_cable/2))-np.log(r_cable/2)  # for shortening expressions
        
        # time parameters
        self.dt = dt
        self.t = [0]
        
        self.m = m  # mass of cargo [kg]
        self.x = [x0]  # list for position values
        self.v = [0]  # list for velocity values
        self.I_list = []  # list for current values
        self.a = [(self.LorentzForce(self.x[0]) + self.Gravity(self.x[0])) /
                  self.m]  # list for acceleration values
```

**Figure 3:** Code snapshot from computational essay on the use of railguns as a mechanism for space elevators

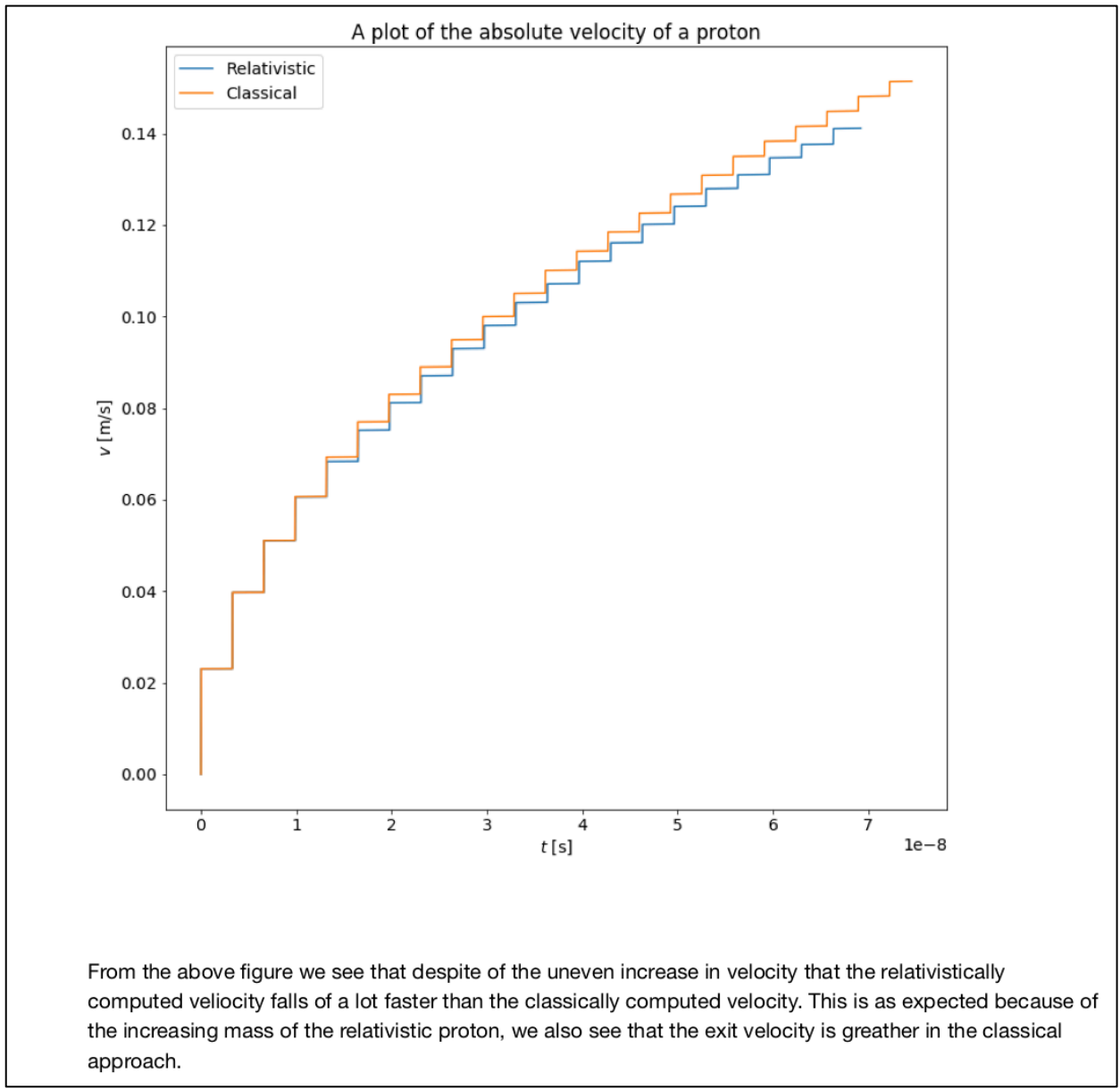

**Figure 4:** Snapshot of the results section from a computational essay comparing non-relativistic and relativistic cyclotron simulations

Beyond the diversity of addressed topics, we also saw a variety of coding approaches on display. Of the 69 essays analyzed (including both the pilot and the full implementation), 10 used higher-order computational techniques, such as student-defined classes, to streamline and clean up the code. Another 33 used student-defined functions for a similar purpose. The remaining 26 students either did not use functions at all, writing their code entirely using a basic set of variables, arrays, and loops, or relied on minimal modifications of the pre-defined functions in the example programs. However, it should be noted that we provided the students with no instructions or guidelines for which coding practices they should use, beyond the expectations described above (i.e., that they needed to substantially build on and modify the example simulations if they chose to use them). So, we cannot take these numbers to be representative of students' coding abilities, only their preferences under the circumstances.

Additionally, we saw several different communicative approaches showcased within the essays. Nearly all students (65/69) used at least one graph or plot to illustrate results, and slightly less than half of the students included an image or diagram (hand drawn or downloaded from the internet) to illustrate parts of their analyses. Students also displayed a variety of narrative structures within their computational essays. Most common was a lab report-type structure, in which the essay acted as a record of what the students had done. These typically featured an introduction section, in which students described their research question; background section, in which they described some relevant theory and equations for their system; analysis section, in which they simulated the system; and a concluding section in which they briefly discussed results and acknowledged limitations.

However, in addition to these report-style essays, we also saw a number of students write computational essays that were framed in other ways, such as stories, mock scientific papers, and personal essays, each of which differed from the lab-report structure in certain ways. *Stories* featured explicit characters, stakes, and plot—for example, one pair of students wrote about the difficulty they had in getting to class on time due to regularly missing the tram in the mornings. One of the authoring students was then written into the essay as a character who used a railgun to launch himself to the lecture hall. *Personal essays* featured a more personal narrative, in which authors guided the reader through their own processes of discovery—for example, one student wrote his essay on the idea of using a hole through the moon (coupled with a railgun) as a mechanism for launching spacecraft outside of the Earth's gravity well, and metacognitively described how this idea developed over the course of the essay. *Mock scientific papers* had a formatting and structure similar to a formal scientific paper, including explicit section headers such as "Introduction", "Background", "Results," "Discussion," and "References", as well as explicit use of scientific terms such as "research question." An example of this is shown in Figure 1. One student, writing such a paper, even included an acknowledgments section which he used to thank the instructional staff for their support. However, it again should be noted that although we provided students with examples, we did not specify how they should structure the written part of their essays beyond the general guidelines given. In future iterations of the project, this is something we intend to develop further.

*4.2 Student reflections on the computational essay writing process*

Although the completed essays provide some insight into student outcomes and experiences, they do not tell the whole story. For this reason, in order to evaluate this design, it was important to talk to students and gain perspective on their processes for writing the essays. In this section, we report on key themes from interviews with students who had recently completed their computational essays. Although we only interviewed a sub-set of students (10 + 12), we found that student responses tended to cluster together into recurring themes. After the first handful of interviews in each semester, we quickly achieved data saturation, where additional interviews no longer provided any novel responses, indicating that our data collection was sufficient. With an eye to helping others replicate our process, in this section we report on themes related to how students used the scaffolds and tools provided and their overall reflections on the computational essay-writing process. In the next section we report on themes related to self-reported learning and attitudinal outcomes from the process. Note that in some cases interviews have been translated from Norwegian into English; in such cases, the translation has been noted after the relevant excerpt.

Students reported spending an average of 10-20 hours on the project, with a few individuals and groups spending significantly less (5-6) and a few significantly more (30+). Students who reported spending a large amount of time on the project often included the time spent pondering possible project ideas and approaches, which tended to significantly boost their time investment. Most students reported heavily using the project description, rubric, and posted examples for guidance. Supporting these reports, we also observed cases of individual essays that either seemed to take inspiration or explicitly re-use code from posted examples. For instance, one of the example essays explored the use of ion drives for sending rockets to Mars; this theme was echoed in a student essay that explored the possibility of using a railgun to send a Tesla Model S car to Mars[2]. Another posted example explored the optimal angle and speed for a volleyball serve when taking into account air resistance, gravity, and the Magnus effect; one pair of students performed a similar analysis when analyzing the optimal mass and angle of a railgun shot in order to maximize distance traveled.

In interviews, we asked students to walk us through the process they went through when writing their essays, and to describe the easiest/most challenging parts of the process. Most students reported that they had found the inspiration for their projects in one of three ways: 1) provided prompts, 2) topics they had read about or seen in popular media, or 3) ideas that spontaneously arose ahead of time or when they started playing around with one of the provided simulations. As an example of students taking inspiration from popular media, Harold and Derek investigated the feasibility of using a railgun to launch supplies from Mercury into orbit around the sun in order to build a Dyson sphere. They described their inspiration as follows:

> **Harold**: *There's this YouTube channel which makes really high-quality videos about different aspects of science, more or less. And one of those videos explored the concept of Dyson spheres. It was just more casually mentioned in that video that a railgun could be a way to do this. But they never went into the math or actually looked at it, more complex than saying a railgun from Jupiter [Mercury] is a possible solution. So, we just took that further.*

This exemplifies a case of students taking inspiration from popular media and trying to explore and model the systems described therein. In contrast, Margaret and Edward also built a model based on the railgun example but took a less direct route to their research question (which was an exploration of the effects of air resistance and the Coriolis force on a long-distance railgun shot):

---

[2] Although we did not interview this set of students, we surmise that they were also inspired by Elon Musk's launch of a Tesla Roadster into orbit around the sun in 2018(38).

> **Edward:** *We didn't really know how to formulate a good question in the start, but we knew we wanted to use the railgun as a model. So, without being too unrealistic we wanted to look at a realistic use of the railgun. We wanted to use the things we learned in mechanics, or Fys-Mek [Introductory mechanics], to model how it would move and what sort of forces would act upon the projectile. So, that's basically what we tried to do.*

Here, Margaret and Edward explain that they found their essay topic by playing around with the railgun simulation, rather than going into the project with a question already defined.

When asked to describe the most difficult part of the essay writing process, interviewees were nearly unanimous: finding an interesting and tractable research question was the most difficult part of the process. For example, Martin put it this way:

> **Martin:** *The most difficult thing for me was figuring out what I wanted to figure out. And figuring out what the project was going to be. That, I think, is always very difficult. (translated)*

We argue that is a positive outcome since it is an authentic part of the research experience (30). That is to say, in professional settings, finding an interesting and tractable research question is often one of the most difficult aspects of a project. Thus, we would argue that this result counts towards our overall project goals of disciplinary authenticity.

Once students had found a viable research question, they described the subsequent physics analysis and coding as being relatively easy and straightforward. In fact, these were often cited as the easiest parts of the project. We interpret this to mean that students, in most cases, were applying their already-existing knowledge of physics and computation to new problems and physical systems, rather than spending excessive amounts of time learning new physics theory or coding practices. To some degree, this might be expected from the project scope—we had primed students to spend roughly 10 hours on the project, which is not sufficient time to learn a great deal of new physics or modeling practices. For this reason, any new knowledge gained about the physics or coding practices was likely to be intuitive or conceptual, rather than technical, as we describe in the next section.

*4.3 Conceptual learning and attitudinal development*

As previously argued, the work produced by the students was novel even if they were likely applying pre-existing skills and knowledge. That is to say, students had not generally worked with the code or phenomena they focused on in their essays prior to this project. So, we argue that the computational essays themselves serve as limited evidence of student learning. However, the completed essays do not allow us to evaluate all of our learning goals for the project. So, in our interviews, we asked students several questions related to conceptual and attitudinal outcomes. Specifically, "is there anything you feel you understand better now than when you started writing your essay," "how do you see computation connecting to physics learning?" and "How would you compare the creative opportunities in this project with the rest of the course?"

With regard to conceptual understanding, most student reported increased understanding of the physical systems they had worked with in their essays, such as cyclotrons or railguns. For example, Emitt and Lily, who simulated particles in the LHC, described an increased understanding of the mechanism behind the LHC:

> **Emitt:** *I feel we underst—at least, I understand the actual structure of the [LHC] complex better, and the way it actually works. Because I knew it was using magnetic—electromagnetism to accelerate the particles, but I didn't know exactly how. Now I know they use the differences in voltages to accelerate it there, and then turn it, and then just do that over and over again. So, I feel it fleshed out the actual—how the thing works, not just that it works.*

Other students reported that they had emerged with a more intuitive understanding of the physical principles used, especially when working with (relatively) abstract concepts like vector fields and 3-dimensional forces. For example, Casey and Gerald reported that they had emerged with a stronger intuition for the Lorentz force through their essay on trapping particles in a magnetic bottle:

> **Casey:** *Definitely, maybe gotten a little bit more intuition about the Lorentz force that we used to find the direction of the force on the particle and how it's affected by the magnetic field. […] I mean, it's not the most complicated force in the world. I mean, it's just a cross-product. But I hadn't really used it in practice, in—practically before we did this, so just visualizing it, seeing the force actually work, making it move. Where we expect the force to move. It's just confirmation. I don't know.*

> **Gerald:** *I think the concept is really cool, where it's sort of like the force is always changing, in towards the center of the bottle. And pushing in and out. And then you can see that in the motion of the particle, as it moves down. Like, in the middle, then the oscillation's bigger and it gets turned further and further in and it gets sort of stopped. It sort of helps the intuition, I guess, to see it like that.*

As Gerald and Casey describe it, the process of working with this force (repeatedly visualizing its effects under various conditions) has given them a deeper intuitive understanding of its meaning.

Other students cited factors like the relationship between different physical quantities—for example, voltage, electric fields, and current. Mary and Cassie pointed to this kind of conceptual development:

> **Mary:** *I feel that I understand the relationship between the electric field, and voltage, and electric current much better. And you get a better understanding of the magnitude of these numbers. Because, I mean, if someone had come and said to me that a lightbulb was 10,000 amps before this project, I would have been like "Oh, yeah, okay" (laughter). But now I know that that wouldn't be realistic. So it was kind of fun to work with the numbers and use them to make sense of their values. (translated)*

> **Cassie:** *Yeah. You had to, kind of, work with it to figure out, "is this here actually realistic?" instead of just accepting their values, sort of. That was kind of cool. In a way, kind of, figure it out yourself. (translated)*

Here, Mary and Cassie report not just a deeper understanding of the relationship between these different quantities but also an intuition for physically reasonable orders of magnitude. Notably, they specify that they would not have developed this intuition had it not been for this project.

Beyond these conceptual outcomes, we would argue that our students developed a more authentic view of the discipline than before they had participated in the project. Partially, this comes from the recurring theme throughout the interviews that students found the process of defining and pursuing a research question to be quite challenging. This, we argue, is much more authentic to the discipline than the more standard class of rote physics problems that students usually spend most of their time solving.

However, several students also explicitly addressed this theme in their interviews, in some cases tying it to the creative affordances of computational essays. For example, Martin directly addressed this topic in his interview, contrasting the computational essay with more standard assignments:

> **Martin:** *So, I do creative things in other parts of my life, but when it comes to studying physics or math, I feel… I take these subjects because they don't require anything from me, in a way, I can just solve the problems and be done with it, and that's comfortable. But the fact that it requires some creativity is—it maybe becomes closer to the way it is to actually do physics. And I feel like this assignment here—that is, it recreates the situation where one has to invent something, one has to find something to figure out, in a way. It's not often that we encounter that in our STEM courses here. So, it's a little bit of a breath of fresh air, creatively speaking. (translated)*

Simon added the following, in his interview:

> **Simon:** *Often, when one is working with standard assignments, you learn a particular method, or you learn a type of problem, or you apply what you've learned to a problem. But here, you're kind of free to try out different paths forward and find your own way. In addition, it's nice, in a way—yeah, at least if one is curious—to dip into other subjects and kind of combine them together. That's what I did with my project, and it was the first time we were able to bring in different areas of physics into the same project. (translated)*

As seen in the transcript excerpts above, students described the process as being creative and motivating in a way that standard physics assignments are not. We argue that this type of scientific creativity is much closer to the process of authentic physics research than solving standard physics problems.

Several students went one step further, and specifically addressed, not just the feeling of authenticity, but the ways in which they could creatively "think their way around" issues that came up during their projects. For example, Rupert and Iris cited this as one of their favorite aspects of the computational essay project:

> **Rupert:** *It's been more enjoyable to work on than other assignments because… I mean. There [with standard assignments], it's "find this, find this, find this, find this." And sometimes, when you get stuck, it gets so that you sit there and just think "Ugh. How do I continue?" While here, it's kind of like, this is what we want to accomplish. So how do we do it? And then it becomes, "now we're stuck, so, ok… this is what we're trying to do." Trying to find a way.*
>
> **Iris:** *So then it becomes a little, like, we can find the way forward that fits us best. Like, when we sat and went back and forth with current versus potential, for example, we could kind of choose which way we wanted to go, choose how complicated it needed to be to do what you wanted. (translated)*

Melanie elaborated on this theme in her interview:

> **Melanie:** *it was nice that if I hit a wall here, I could just do something else. In a normal assignment, if I hit a wall, I have to, like, knock through it.*

We argue that these excerpts are notable because the students seem to be pointing to a heretofore unmet need in the physics curriculum—that is, these students appreciated the opportunity to explore challenging problems in which there was not just one right answer, which allowed for the possibility of alterative paths to a solution. This, we argue, is one of the primary strengths of the computational essay design, and another way in which it mimics professional physics practice.

**4. Discussion and Implications**

As discussed, one of our primary motivations in this project was to help our students develop a more authentic view of the physics discipline. Upon reflection, we are encouraged by both the quality of essays produced and the student responses described above. We consider both of these to be markers of success—the wide variety of essay topics shows that students took seriously the challenge of defining and pursuing a research question, and the interview responses show that students were also aware (at least upon reflection) of the differences between this kind of work and the more standard assignments in the physics curriculum.

We argue that the computational essay scaffolds were an essential part of this success. Just as important, however, were the things we chose to avoid or leave out. For example, we deliberately provided students with examples that they could model their projects off of, but did not provide them with templates to fill in (as is sometimes done with scaffolded lab reports). This ensured that students were free to experiment with different styles of questions and written narratives, rather than being constrained to one genre. We also did not specify in advance which types of questions students should seek to answer, or which physics they should use to do it.

There are certainly limitations to this design. First, we did not check or evaluate computational essays for correctness, and in fact we saw some cases in which students used clearly-incorrect physics, especially when it came to energy conservation within magnetic fields. However, since our students had little experience with open-ended projects prior to this semester, we deliberately chose to err on the side leniency and extra support, rather than being overly critical. Our philosophy is that incorrect results are not bad in and of themselves—mistakes are part of the scientific process (31), and one useful aspect of computational essays is that they open up for the possibility of having students revise their work in response to critiques. In the current design, however, we did not build in any room for such critique, and we plan to incorporate this into future versions of the design.

Relatedly, students received little feedback on their computational essays. Most students received one or two questions after their presentations, and essays were graded (using the rubric) by a TA, but one recurring point of feedback from students was that they wished they had received more feedback on their work. One option for addressing this limitation is having students peer-review each other's work, and we are currently in the process of exploring this idea in future iterations of the design. This solution would have the added benefit of also addressing the first limitation, and would additionally add another layer of scientific authenticity.

We also noted some limitations in our rubric-based assessment. In most cases, students fulfilled or exceeded the defined criteria; however, in a few cases students were able to "slip through" our assessment criteria by "gaming" the rubric we provided. For example, we noticed a handful of groups who focused a great deal on the research question and written presentation (achieving full points on those parts of the rubric) while making minimal modifications to their code. Although these students passed, we feel that they missed one of the main objectives the project, the experience of developing a computational simulation. In interviews, many of these students admitted to feeling poorly prepared or deeply uncomfortable with the coding parts of the course. In future iterations, we intend to explore how we can provide added support for such students.

Finally, we note that this intervention presupposed a certain level of pre-existing physics computational literacy (32) on the part of our students. Universities that have not integrated computation into their physics courses to the same degree as the University of Oslo may struggle to replicated this intervention. However, we would argue that computational essays, as a class of assignment, are sufficiently flexible that they could be adapted to a variety of circumstances, from courses that provide a first-exposure to scientific programming (33) to advanced computational physics courses. The scaffolds we used can also be tuned and adapted to provide extra support, if needed, by giving students additional guidelines and questions to investigate, or adjusting the complexity of the example simulations.

Despite these limitations, we see this project as a significant step forward for our curriculum. Our goal with the project was to provide students with the experience of doing open-ended, authentic, computational physics investigations. This goal seems to have been met. In the longer term, our hope is that experiences like these may give students more agency over their own learning(8) and a better sense of what it means to do physics research. This, in turn, may help us to retain those physics students who are frustrated with the highly-structured nature of standard undergraduate physics education and who might therefore be at risk of not finishing their studies or transferring to a different program.

We see several implications for this project. First, this kind of work may help students to develop physics computational literacy (32). As discussed, computational essays provide students with a way to showcase their coding skills, their modeling skills, and their communication skills. However, they also provide an opportunity to synthesize these skills together. This kind of synthetic and creative work, which operates at a higher level of thinking on scales like Bloom's Taxonomy (34), can help students develop skills and knowledge that might otherwise go unaddressed within a standard physics curriculum.

Second, we propose that this project can set the foundation for a larger-scale learning progression focused on developing students' computational and inquiry-based skills. In future courses, once students are comfortable with the expectations and workflow for these kinds of projects, we can lay on additional criteria such as explicit requirements that they use higher-order computational methods, writing guidelines, and/or revisions their work in response to critiques. We see great potential for such a progression in preparing students to do longer, more detailed, and complex projects near the end of their bachelor's degree and/or smoothing the transition into a master's project. Computational essays might also be used as a form of assessment, either in a limited way (for example, scaffolded take-home exam problems), or through a portfolio-style structure in which students write a series of computational essays throughout their bachelor degrees, which can then be shown to future employers. We have found that computational essays are easily publishable and shareable online, and digital solutions like our computational essay showroom (29) could easily be used to store and publicize student portfolios.

Third, we argue that this computational essay design provides an alternative approach for having students do authentic, open-ended, inquiry-based, research-like work that does not require any additional laboratory equipment (18,35). This alternative could be especially useful to universities that have an ambition to move their physics instruction in this direction but cannot afford the expensive equipment necessary to facilitate this kind of work in a laboratory setting. As previously argued, computation is just as authentic to modern-day physics practice as laboratory work (and is also very appealing to employers in a variety of industries), so we suggest that universities would lose little in adopting this approach.

Finally, we propose that this kind of project could also be used in interdisciplinary settings such as multidisciplinary data science (33), bioinformatics, or computational humanities (36) courses, thereby facilitating connections between physics and other disciplines. Computation is a shared language across scientific and social-science disciplines, and computational essays like those described above can easily serve as tool for interdisciplinary collaboration.

**5. Conclusion**

In this paper, we report on initial results from an educational design implementing computational essays in undergraduate physics. We have set out to provide guidelines in how to run such a design, as well as a taste of the results. Our initial outcomes were positive—not only did we see evidence of conceptual and attitudinal development in our students, but nearly all students we interviewed recommended we continue with the design in future semesters. Our intention in reporting these methods and results is that other researchers, educators, and curriculum designers will adopt and use computational essays in their own programs—we see a great need to educate future physicists in this emerging practice for performing and communicating research.

In the future, we intend to continue developing this design. We plan to tighten the guidelines to address the limitations described above and expand the set of examples and supports provided to students. We intend to use computational essays in additional courses, especially those later in the physics degree program, and build on this established foundation. We also intend to further develop the communicative aspects of computational essays, educating students in the goals and expectations for this new genre of writing.

Overall, we see great promise in the educational possibilities of computational essays, and hope they will be adopted far and wide.

## 6. Acknowledgments

This project was funded by Norges Forskningsråd Project Number 288125 "International Partnership for Computing in Science Education," and NOKUT (The Norwegian Agency for Quality Assurance in Education) which supports the Center for Computing in Science Education. We would like to thank Danny Caballero, Elise Lockwood, Crina Damsa, Christine Lindstrøm, John Burk, and Karl Henrik Fredly for their feedback and help with this study.